# LogMin: A Model For Call Log Mining in Mobile Devices

K.S. Kuppusamy[1], Leena Mary Francis[2], G. Aghila[3]

[1,3]Department of Computer Science, School of Engineering and Technology,
Pondicherry University, Pondicherry, India
[1]`kskuppu@gmail.com`, [2]`rosebeauty02@gmail.com`, [3]`aghilaa@gmail.com`

## ABSTRACT

*In today's instant communication era, mobile phones play an important role in the efficient communication with respect to both individual and official communication strata. With the drastic explosion in the quantity of calls received and made, there is a need for analyses of patterns in these call logs to assist the user of the mobile device in the optimal utilization. This paper proposes a model termed "LogMin" (Log Mining of Calls in Mobile devices) which is aimed towards mining of call log in mobile phones to discover patterns and keep the user informed about the trends in the log. The logging of calls would facilitate the user to get an insight into patterns based on the six different parameters identified by the proposed LogMin model. The proposed model is validated with a prototype implementation in the Android platform and various experiments were conducted on it. The results of the experiments in the LogMin Android implementation validate the efficiency of the proposed model with respect to user's relevancy metric which is computed as 96.52%.*

## 1. INTRODUCTION

Mobile phones have become the "numero uno" devices not only among the nerds but also with general masses which is validated with the terrific explosion in the quantity of mobile devices. The availability of more than 6.8 billion mobile devices across the globe emphasises the utility and the necessity of these devices across the spectrum of people. The drastic increase in the number of mobile devices is substantiated by the fact that the quantity of devices in 2010 is 5.4 billion and 6.1 billion in 2011. [1] With respect to the Indian context the total number of mobile phone subscription is about 906 million amounting to 73% of the population which is in line with the global average.

In addition to the drastic increase in the quantity of the mobile devices, their individual utility is also on a steep rise. The utility of mobile devices among the youth in specific, itself has been studied by researchers in various dimensions including the psychological analysis. [2], [3] As with any other technology, the mobile phones do carry some adverse effects like behavioural addiction. In order to handle these problems the study of mining the call log of mobile phones becomes an interesting and necessary research issue which need to be addressed.

The general utility of mobile phones among the end users in terms of various aspects like communication pattern, management of activities etc have been studied by researchers to facilitate efficient handling of these mobile devices.[4].

This paper proposes a model titled "LogMin" (Log Mining of calls in mobile devices). The objectives of the proposed LogMin model are as listed below:

- Proposing a model for mining the log of calls in the mobile devices to infer the user relevant information from them.
- Providing a detailed insight into the call logs of the mobile phones to the users, in terms of multiple components like Time-of-Day, Provider based grouping with the help of clustering techniques.

The remainder of this paper is organized as follows: In Section 2, some of the related works carried out in this domain are explored. Section 3 deals with the proposed model's mathematical representation. Section 4 is about prototype implementation and experiments. Section 5 focuses on the conclusions and future directions for this research work.

## 2. RELATED WORKS

This section walks though some of the related works which have been carried out in this domain. The proposed LogMin model attempts to explore the patterns in the call log and provide the user various views along different dimensions. The mobile phones and their utility has been an important research area in the information communication domain which has been studied by various researchers to address specific issues.

The mobile phone logs have been studied by researchers for various tasks like clustering of users and interface optimization purpose [5]. In addition to the logs, various other sensing abilities from the mobile phones have been utilized by researchers in understanding the optimal utilization of these devices [6] [7].

With respect to handling the call log of mobile phone, there are studies which have focused on providing calendar specific categorization of logs [8]. There exist applications for various smartphone platforms like Android for managing the call log in a professional manner which provide facilities like extraction of log information in a specific format, providing the statistical data regarding the log in the mobile devices [9]. These kind of applications provide various services like rendering of log in specific manners apart from the traditional linear list manner, facility to backup and restore the log information and the ability to manipulate the log information at the lowest possible abstraction [10]. There are studies on reality mining based on the real time datasets which considers various log information provided by the mobile devices [11]. In addition to the individual log mining there are studies conducted at the corporate level to analyse the call log of technical support etc [12]. There are studies conducted on the call logs from central repositories with millions of calls [13]. The call pattern analysis is carried out using neural networks by researchers [14].

The proposed LogMin model provides the call log results in the form of clusters based on parameters like duration etc. There clustering techniques have been studied in detail in the information management domain by various researchers [15], [16], [17]. The clustering techniques have been adopted for various applications like image handling, web page manipulation etc [18]. The proposed model utilizes the k-Means clustering technique for the purpose of clustering the call logs into various groups based on different parameters. The k-Means clustering algorithm allows to cluster a dataset into k-clusters and it belongs to the supervised machine learning category [19], [20]. The reasons for selecting the k-Means algorithm are the ability to cluster faster than other techniques like hierarchical, fuzzy clustering and the ability to produce the clusters in a cohesive manner.

International Journal in Foundations of Computer Science & Technology (IJFCST), Vol. 3, No.4, July 2013

The proposed LogMin model extracts the call log information from the mobile devices and the performs the analysis on the log data based on various parameters and provide the clustered result set with the help of k-Means clustering technique. The details on the model are illustrated in the following section.

## 3. THE LOGMIN MODEL

This section explores about the building blocks of proposed LogMin model for analysing the call log of the mobile phones and inferring details from it. The block diagram of the proposed LogMin model is as shown in Fig. 1.

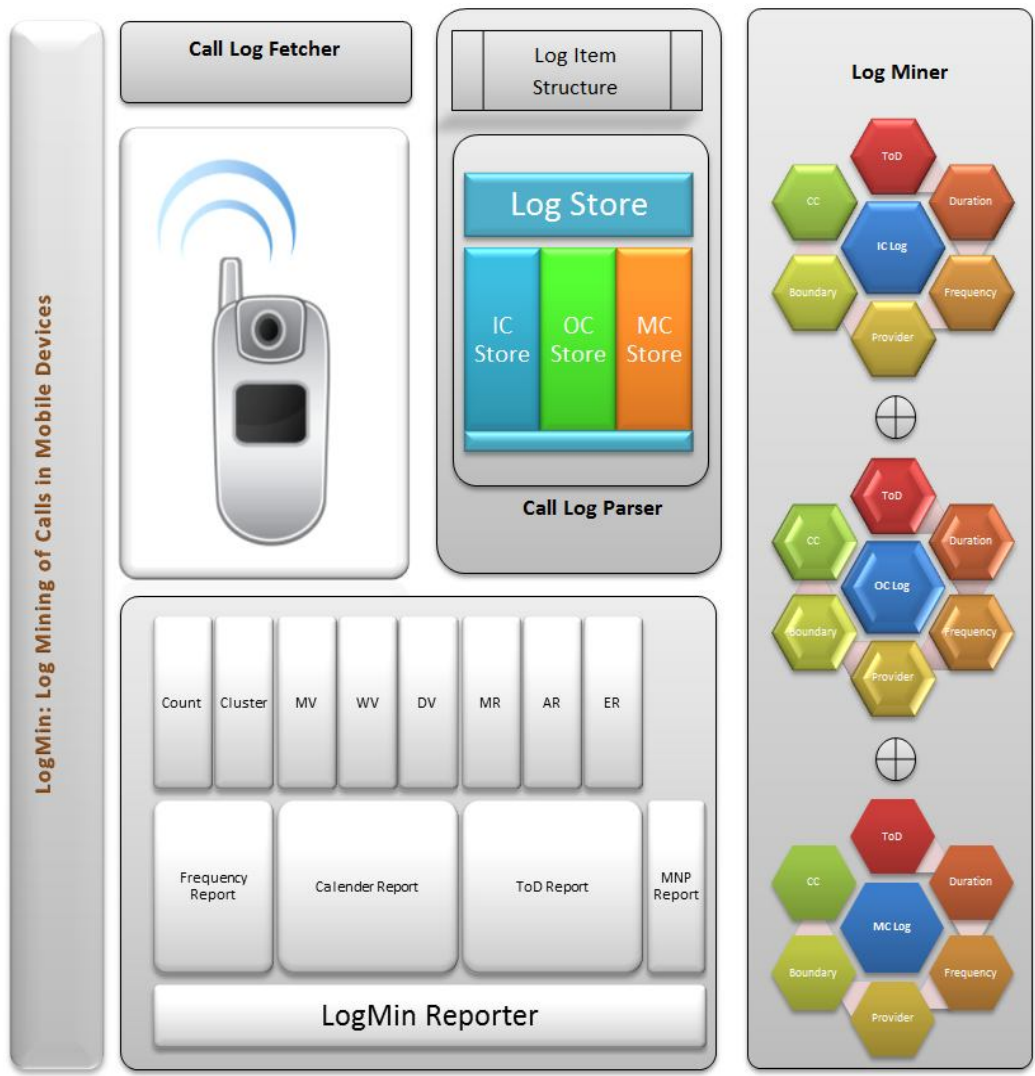

Figure 1. LogMin – Block Diagram

The components of the proposed LogMin model are as explained below:

- **Call Log Fetcher**: The role of Call Log Fetcher is to fetch call log information from the device's call repository.

- **Call Log Parser**: The Call Log Parser receives the input from the call log fetcher and its primary aim is to parse the input and store it appropriate locations. It has got the following sub components.

    o **Log Item Structure**: The log item structure is a data structure used to hold component data of the call log. This structure has following fields: The "Call Log Id" which is a unique number assigned to every single entry in the log in order to identify a call item uniquely. This is assigned by the model automatically in an auto-increment manner. The "Call Number" is the phone number of the device from which the call has originated. In case of outgoing calls it would hold the value of "SELF" which indicates that the caller is the current device. In case of items in the contact there would be a "Caller Name" field. In case of new number this field would be populated as "Un-Known". The next field in the structure is the "Start-Time" of the Call and the last field is the "End-Time" of the call. This structure's visual representation is as shown in Figure 2.

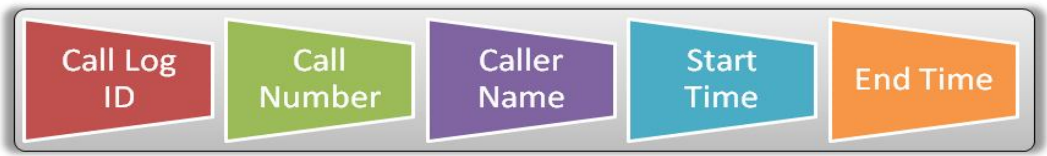

Figure 2: Log Item Structure

    o **Log Store:** Another sub-component in the Log parser is the "Log Store" which has the facility to store the calls into three different sectors which are termed as "Incoming (IC) Store", "Outgoing (OC) Store" and "Missed Call (MC) Store". The missed calls are identified by comparing the start time and end time in the call log structure. In cases where they are equal that particular call is identified as the missed call.

- **Log Miner**: The primary role of Log Miner is to mine the logs in all the three different stores explained above. The mining of the calls is carried out using six different parameters which are identified as "ToD – Time of the Day", "Duration", "Frequency", "Provider", "Boundary", "CC -Conference Count". The boundary parameter indicates the calls which the user has either made or received in both sides of the time dimension. The "Conference Count" is another parameter which indicates the parallel calls i.e. the conference calls made with the current call. The concurrent calls are identified with the the help of time window which is obtained by comparing the start time and end time of these calls.

- **LogMin Reporter**: The LogMin reporter is component is responsible for providing the reports based on the results of Mining component. The various reports which are presented to the user are Frequency report, calendar report, ToD (Time of Day) report and "MNP – Mobile Number Portability report". These reports are generated by receiving the inputs from the miner component. The frequency report is provided in two dimensions viz., Count and Cluster. The cluster report is generated by clustering the calls based on the frequency which is done with the help of k-Means clustering algorithm. The calendar report is presented in three dimensions viz., "MV – Month View", "WV – Week View" and "DV Day View". Another report presented to the user is the "TOD report" which is again rendered in three different variations viz., "MR- Morning Report", "AR – Afternoon report" and "ER – Evening Report". These are based on the Time of Day at which the calls are either received or made. Another report generated by the LogMin reporter is the mobile number portability report which is based on optimizing the tariff to the user based the call log i.e. if the user is making frequent calls numbers belonging to the a particular provider other than the current provider then the decision to port the number to that service provider is analysed. The portability index is computed and if it goes beyond the threshold value then MNP is advised by the LogMin.

### 3.1 Mathematical Model

This section explores the mathematical representation of the proposed LogMin model. The LogMin call log store with its three components is represented as shown in (1).

$$\Omega = \begin{Bmatrix} \alpha \\ \beta \\ \chi \end{Bmatrix} \quad (1)$$

In (1), $\Omega$ indicates the Call Log Store, $\alpha$ indicates the IC Store, $\beta$ indicates the OC Store and $\chi$ represents the MC Store. The Log Structure is represented as shown in (2).

$$\Gamma_G = \begin{Bmatrix} \delta \\ \phi \\ \varphi \\ \gamma \\ \eta \end{Bmatrix} \quad (2)$$

In (2), $\Gamma_G$ is the generic Call Log Structure, and the five components indicated in Figure 2 are represented from $\delta$ to $\eta$ respectively.

The generic $\Gamma_G$ is extended to all the three specific log stores as shown in (3).

$$\Gamma = \begin{Bmatrix} \Gamma_{IC} \\ \Gamma_{OC} \\ \Gamma_{MC} \end{Bmatrix} \quad (3)$$

The entries in all these stores are analysed using six different dimensions as indicated in (4).

$$\alpha_M = \begin{Bmatrix} \mu \\ \nu \\ o \\ \pi \\ \varpi \\ \theta \end{Bmatrix} \quad (4)$$

In (4), $\alpha_M$ indicates the mining of IC Store and six different parameters utilized for the purpose of mining are represented from $\mu$ to $\theta$ respectively. The parameters are computed individually. The computation of ToD parameter $\mu$ is as shown in (5). The ToD is computed as the mean of start time and end time which are represented as $\vartheta$ and $\rho$ respectively.

$$\alpha_M = \begin{Bmatrix} \mu \to \dfrac{\vartheta + \rho}{2} \\ \nu \\ o \\ \pi \\ \varpi \\ \theta \end{Bmatrix} \quad (5)$$

The duration parameter $\nu$ is computed as shown in (6). The duration is computed as the difference between the start and end time.

$$\alpha_M = \begin{Bmatrix} \mu \to \dfrac{\vartheta + \rho}{2} \\ \nu \to |\rho - \vartheta| \\ o \\ \pi \\ \varpi \\ \theta \end{Bmatrix} \quad (6)$$

The frequency parameter is computed as the count of calls in the reference time as shown in (7).

International Journal in Foundations of Computer Science & Technology (IJFCST), Vol. 3, No.4, July 2013

$$\alpha_M = \begin{Bmatrix} \mu \to \dfrac{\vartheta + \rho}{2} \\ \nu \to |\rho - \vartheta| \\ o \to |\forall \Gamma(\phi) : \Gamma(\gamma) > T_R| \\ \pi \\ \varpi \\ \theta \end{Bmatrix} \quad (7)$$

In (7), $T_R$ indicates the Reference Time after which all the computation is made. The reference time is incorporated to avoid very old calls skewing the results. The Provider data is retrieved from the online repositories as shown in (8).

$$\alpha_M = \begin{Bmatrix} \mu \to \dfrac{\vartheta + \rho}{2} \\ \nu \to |\rho - \vartheta| \\ o \to |\forall \Gamma(\phi) : \Gamma(\gamma) > T_R| \\ \pi \to P(\Gamma(\phi)) \\ \varpi \\ \theta \end{Bmatrix} \quad (8)$$

The boundary parameter is computed as the set of entries in the call log before and after the call with in a reference time, as shown in (9).

$$\alpha_M = \begin{Bmatrix} \mu \to \dfrac{\vartheta + \rho}{2} \\ \nu \to |\rho - \vartheta| \\ o \to |\forall \Gamma(\phi) : \Gamma(\gamma) > T_R| \\ \pi \to P(\Gamma(\phi)) \\ \varpi \to |\forall \Gamma(\phi) : T_F \geq \Gamma(\phi(\gamma)) \geq T_P| \\ \theta \end{Bmatrix} \quad (9)$$

In (9), $T_F$ indicates the Reference time for calls after the current one, and $T_P$ indicates the Reference time for calls before the current one. The conference count parameter is computed by analysing the start and end time of calls and their proximity, as the conference calls would have near equal, start time and end time. The computation is as shown in (10).

$$\alpha_M = \begin{cases} \mu \to \dfrac{\vartheta + \rho}{2} \\ \nu \to |\rho - \vartheta| \\ o \to |\forall \Gamma(\phi) : \Gamma(\gamma) > T_R| \\ \pi \to P(\Gamma(\phi)) \\ \varpi \to |\forall \Gamma(\phi) : T_F \geq \Gamma(\phi(\gamma)) \geq T_P| \\ \theta \to |\forall \Gamma(\phi) : \vartheta \approx \Gamma(\gamma); \rho \approx \Gamma(\eta)| \end{cases} \quad (10)$$

The same procedure is repeated with corresponding data for $\beta_M$ and $\chi_M$. After the completion of parameter computation, the LogMin miner analyses the data with statistical techniques for some of the reports and it utilizes the k-Means clustering technique for computing the clusters in the log data. The clustering process is as indicated in (11).

$$LM_C = \begin{cases} \forall \Gamma_i & \Re(\Gamma_i) \to \{\omega_1, \omega_2 .. \omega_k\} \\ & i = [0, |\Gamma|] \end{cases} \quad (11)$$

In (11), $LM_C$ indicates LogMin Cluster, $\Re$ indicates the k-Means clustering engine, and each $\omega_i$ represents a cluster, and "k" indicates the cluster size. The result of this process is given to the LogMin reporter which generates the reports as per the user's request.

## 4. EXPERIMENTS AND RESULT ANALYSIS

This section explores the experimentation and results associated with the proposed LogMin model for mining the call log of mobile devices. The prototype implementation is done with the Android platform [21]. The prototype application is tested with an array of hardware running Android operating system. Though the prototype is made with the Android platform, the same shall be extended to other environments like iOS and Windows Mobile. The screenshot of the application is shown Figure 3.

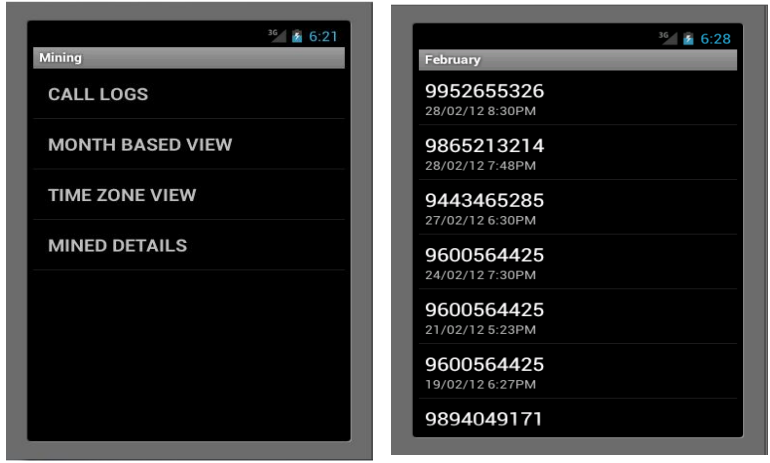

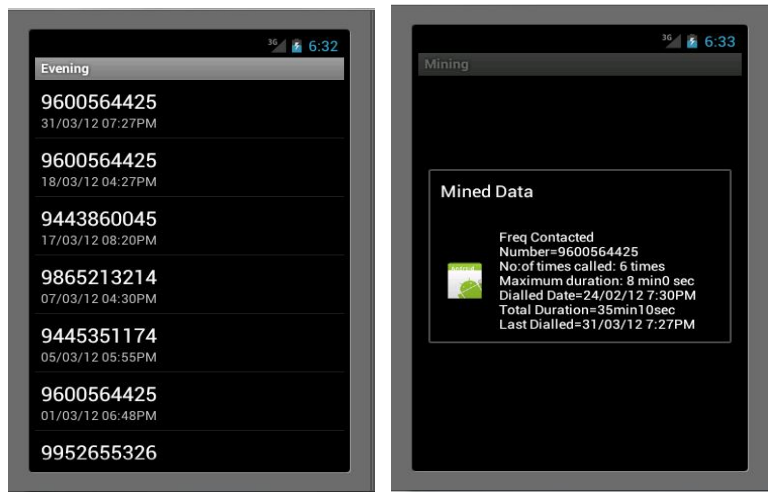

Figure 3: Screenshots of LogMin Android Application

The experiments were conducted with 15 groups of users covering a spectrum from Novice, General to Geek users. The relevance of the data mined with the LogMin model was analysed with these groups of users and the experiments were conducted in various sessions. The results of the experiments are illustrated in Figure 4 and Table 1. During the experiments the users were asked to mark the information rendered by the LogMin reporter into three different categories viz., "CRI – Completely Relevant Information", "PRI – Partially Relevant Information" and "CII – Completely Irrelevant Information".

Table 1: LogMin Information Relevance

| Session ID | CRI | PRI | CII |
| --- | --- | --- | --- |
| 1 | 88.4 | 6.4 | 5.2 |
| 2 | 82.6 | 10.3 | 7.1 |
| 3 | 85.5 | 12.8 | 1.7 |
| 4 | 86.2 | 12.5 | 1.3 |
| 5 | 88.1 | 11.4 | 0.5 |
| 6 | 80.1 | 14.5 | 5.4 |
| 7 | 82.1 | 10.5 | 7.4 |
| 8 | 88.5 | 6.5 | 5 |
| 9 | 90.2 | 6.5 | 3.3 |
| 10 | 92.3 | 5.5 | 2.2 |
| 11 | 95.1 | 3.2 | 1.7 |
| 12 | 94.2 | 4.5 | 1.3 |
| 13 | 91.1 | 5.9 | 3 |
| 14 | 90.1 | 6.5 | 3.4 |
| 15 | 89.8 | 6.5 | 3.7 |

International Journal in Foundations of Computer Science & Technology (IJFCST), Vol. 3, No.4, July 2013

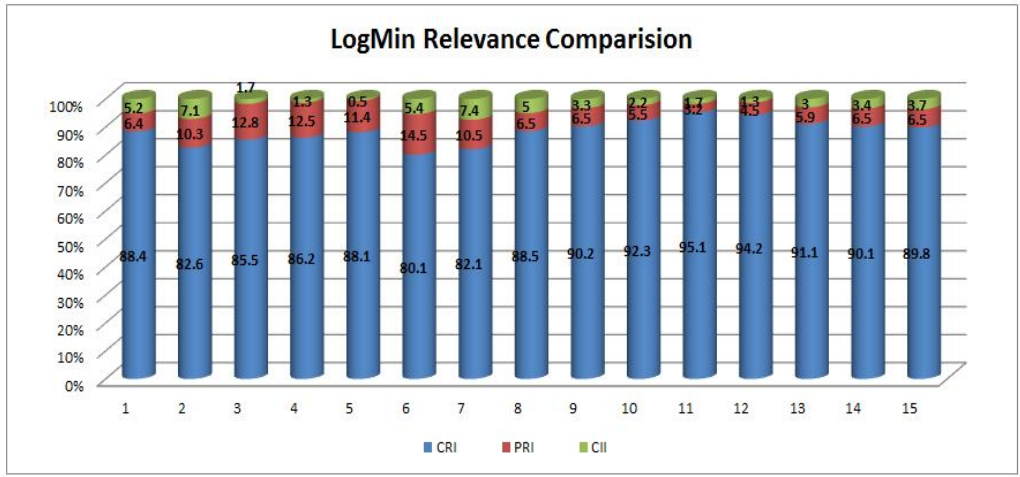

Figure 4: LogMin Relevance Comparison

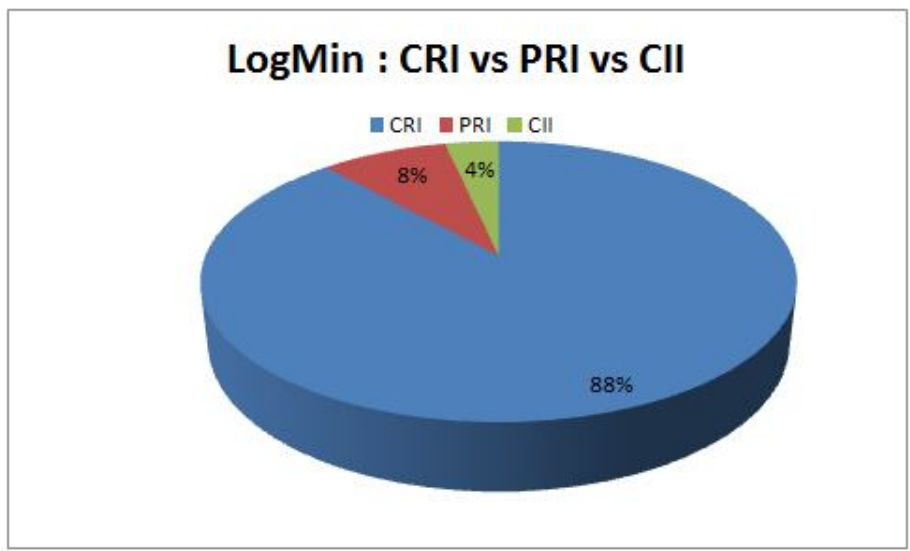

Figure 5: LogMin CRI vs PRI vs CII

It can be observed from the experimental results that the mean of CRI across the sessions is 88.2%, the mean of PRI as 8.23%. The CII component is observed only at a negligible level of 3.48% as illustrated in Figure 5. As the cumulative of CRI and PRI is evaluated as 96.52% the efficiency of the LogMin model in providing the relevant information from the call log is confirmed.

## 5. CONCLUSIONS AND FUTURE DIRECTIONS

The following are the conclusions derived from the LogMin model:

- LogMin model provides a detailed insight into the calling behaviour of the mobile user which shall be used to improve the effectiveness of utilization.
- The proposed LogMin model facilitates the retrieval of user relevant information from the logs of calls from the mobile device.
- The efficiency of the LogMin model is confirmed with the user relevance metric with a cumulative PRI and CRI value of 96.52%.

The future directions for the LogMin model include the following:

- The LogMin model shall be extended to include the log information other than the call log like GPS and other sensor data.
- The model shall be further enriched by incorporating collaborative call log mining which would include the log of users who communicate with each other frequently, thereby forming call log based network.

International Journal in Foundations of Computer Science & Technology (IJFCST), Vol. 3, No.4, July 2013

**Authors**

Dr. K.S.Kuppusamy is an Assistant Professor at Department of Computer Science, School of Engineering and Technology, Pondicherry University, Pondicherry, India. He has obtained his Ph.D in Computer Science and Engineering from Pondicherry Central University, and the Master degree in Computer Science and Information Technology from Madurai Kamaraj University. His research interest includes Web Search Engines, Semantic Web and Mobile Computing. He has made more than 20 peer reviewed international publications. He is in the Editorial board of three International Peer Reviewed Journals.
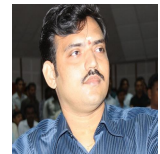

Leena Mary Francis was associated with Oracle as Software Engineer before she joined as Assistant Professor at Department of Computer Science, SS College, Pondicherry, India. She has obtained her Master's degree in Computer Applications from Pondicherry University. Her area of interest includes Web 2.0 and Information retrieval.
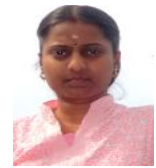

Dr. G. Aghila is a Professor at Department of Computer Science, School of Engineering and Technology, Pondicherry University, Pondicherry, India. She has got a total of 24 years of teaching experience. She has received her M.E (Computer Science and Engineering) and Ph.D. from Anna University, Chennai, India. She has published more than 75 research papers in web crawlers, ontology based information retrieval. She is currently a supervisor guiding 8 Ph.D. scholars. She was in receipt of Schrneiger award. She is an expert in ontology development. Her area of interest includes Intelligent Information Management, artificial intelligence, text mining and semantic web technologies.
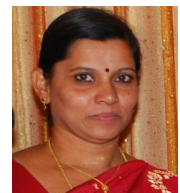